# Mobile Advertisement in Vehicular Ad-Hoc Networks

Ciprian Dobre and George Cristian Tudor
Department of Computer Science
University POLITEHNICA of Bucharest
Spl. Independentei, 313, Bucharest
Romania
E-mails: ciprian.dobre@cs.pub.ro, george.tudor@cti.pub.ro

**ABSTRACT**

Mobile Advertisement is a location-aware dissemination solution built on top of a vehicular ad-hoc network. We envision a network of WiFi access points that dynamically disseminate data to clients running on the car's smart device. The approach can be considered an alternative to the static advertisement billboards and can be useful to business companies wanting to dynamically advertise their products and offers to people driving their car. The clients can subscribe to information based on specific topics. We present design solutions that use access points as emitters for transmitting messages to wireless-enabled devices equipped on vehicles. We also present implementation details for the evaluation of the proposed solution using a simulator designed for VANET application. The results show that the application can be used for transferring a significant amount of data even under difficult conditions, such as when cars are moving at increased speeds, or the congested Wi-Fi network causes significant packet loss.

## 1. INTRODUCTION

Intelligent Transport Systems (ITS) are formed by adding information and wireless communication capabilities to transport infrastructure and vehicles. Such systems have the potential to increase safety and comfort of drivers (Tarnoff, *et al*, 2009). ITS applications can lead to cooperative collision warning, to congestion-avoidance and the finding of faster and safer routes for drivers. A different category is represented by the non-safety (comfort) applications. Examples include Electronic Toll Collection (ETC) (Lee, *et al*, 2008), car to home communications (Dkesson&Nilsson, 2002), travel and tourism information distribution (O'Grady&O'Hare, 2004), etc.

In this paper we present a solution for the dissemination of information to interested drivers using ITS capabilities. *Mobile advertisement over VANET* is an application that takes advantage of short-range wireless network communication for making recommendations based on location-awareness. It uses Access Points (AP) as emitters, constantly sending messages to all vehicles within range. The main advantage of using short-range wireless technology for this application is that it can exploit the natural locality of the signal. The information is received by vehicles located within the neighboring area. Therefore, by using a short-range signal, the difficult task of finding vehicles within a geographical region is avoided.

Using WiFi communication is difficult because of the cars' speeds. Cars enter and leave the wireless transmission range of an AP at all times, so the window for transmitting the data is low. However, by directly incorporating data into the WiFi beacons we manage to solve this. We show results proving the capability of such an approach to cope with increased car's speeds.

The messages range from commercial advertisements to information about road conditions or traffic congestion. The client application runs on the driver's smart phone, and it receives messages encapsulated in the beacon data sent over the wireless protocols by neighboring access points. Based on the driver's specific preferences, the client is capable to filter and recommend specific information to the end-user.

Such an application has the potential to replace the static and more expensive advertisement traditional billboard panels. There are advantages for both the driver, as well as the business provider. The driver is presented with less information, which is more relevant to its specific needs. The business provider advertises its products in a more efficient way. Typical scenarios include a restaurant dynamically advertising its menu or a store advertising its discounts. Or the administrator of a road decides to do some maintenance work on a specific portion. In this case he/she could place APs at locations leading up to the affected area, informing drivers of the hardened driving conditions and possibly offering advice about alternative routes.

In this paper we present the architecture and implementation details of the proposed solution. We present details describing the use of APs for mobile advertising. We outline the obstacles on the implementation and proof of the theoretical expected performance. We also present evaluation results of the proposed solution using modeling and simulation. For that we used a simulator equipped with a realistic mobility model, with streets, routes and various driver behaviors, as well as having a network model including wireless communication protocols adequate for the use in vehicle ad-hoc networks.

The rest of the paper is structured as follows. Section 2 presents the solutions for mobile advertisement in vehicular ad-hoc networks. In Section 3 we present implementation details of a prototype solution. In Section 4 we present experimental results that demonstrate the capabilities of the presented solution. Finally, in Section 5 we give conclusions and present future work.

## 2. MOBILE ADVERTISING USING ACCESS POINTS

The use of access points for implementing mobile advertisement is appealing because the wireless signal is confined to a specific area. In this case the information is transmitted only to vehicles located in the same geographic region with the AP. In many cases the information is of interest only to vehicles situated in the vicinity of the data source. The information might be, for example, the advertisement of a local sale or a special discount in a restaurant. The solution does not require a discovery phase, and, therefore, it reduces the time needed for the propagation of the information.

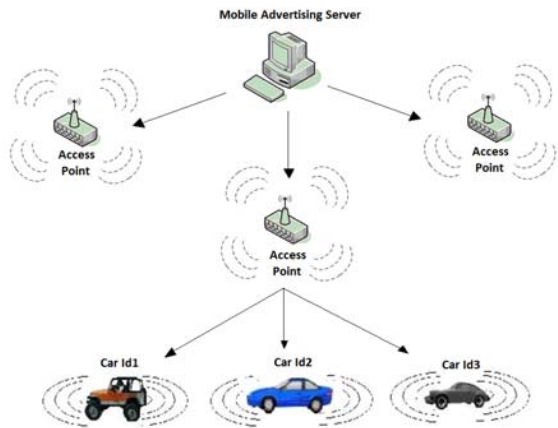

Figures 1: System architecture.

In addition, we wanted to develop a solution that can work in a mobile dynamic environment formed between the AP and vehicles constantly changing their positions. According to the 802.11 standard (which regulates the use of wireless local area networks), the nodes must establish a connection prior to any actual communication (Wong, *et al*, 2006). In case of the proposed mobile dissemination solution this is not a viable solution. In this case vehicles are moving at relatively high speeds (50-60 km/h in a typical urban scenario) and, considering that most APs have a limited range of transmission (around 100 m), the cars would not stay in the communication area long enough for a connection to be established, let alone send the required information. Also, a standard AP can only support a limited number of connections at one time, thus causing problems in case of dens traffic. Vehicles would have to send packets back to the access point, thus leading to possible packet collisions and increased packet loss ratio. Any vehicle could only receive messages from one AP at a time.

Because establishing a connection between nodes is not a viable solution, we had to work around the limitations of the 802.11g standard. We propose using beacon frames for carrying the data. According to the wireless protocol, such beacon frames are periodically sent by APs to advertise their presence over the network. These frames can be modified to carry additional information besides the regular network control parameters.

Figure 1 illustrates the general design of the mobile advertising system. Messages to be broadcasted are managed by a mobile advertisement server, which can be connected to several access points. The server passes the messages to the APs, which broadcast them over the wireless network. Vehicles within each AP's wireless transmission range, equipped with wireless devices, receive and filter messages. Based on the user's preferences, a mobile client can further discard or present the message to the driver. For beacon collision avoidance we consider solutions described in (Sgora, *et al*, 2009).

The business provider can dynamically change the advertisement message sent by each individual AP. It can also dynamically manage operational parameters such as the beacon interval, or the advertised wireless network name. Modifying the beacon interval is useful for creating an energy efficient system. As presented in the next Section, data throughput is proportional with the interval between beacon frames. A smaller interval can result in more data being transmitted. For power saving the advertising server can set the beacon interval to a higher or smaller value, depending on the size of the message being transmitted.

The name of the wireless network is used by cars to identify the originator of the received message. In standard wireless networks this parameter should uniquely identify an AP and is set by default to the MAC value. For our solution we took advantage of this field to increase the system's performance. If more APs share the same network name a vehicle passing within range could not differentiate between them. To the vehicle it would appear as if it receives a constant flow of information from a single source. Considering that the wireless range of an access point is limited to about 100 meters, this technique can be useful for increasing the area over which a message is transmitted. A vehicle can seemingly roam from one AP to another without knowing that the source of the message changed by placing multiple APs in such a way that no gap in the wireless signal exists between them.

The solution is based on a push model of information delivery. The emitter keeps sending data without expecting a response. The idea is to overload the 802.11g beacon frames with additional information that can be interpreted and used by the receivers. The beacon frames are sent by APs at fixed intervals to make their presence known over a Wi-Fi network. These frames are received by all clients in range, regardless whether they are connected or not to the AP. A client can also simultaneously receive beacon frames from multiple APs. For that it periodically scans all channels to receive incoming messages encapsulated within beacons.

The custom information is added to a beacon frame by altering its structure and overriding one or more fields. There are *three fields* that can be used (Wong, *et al*, 2006). The **SSID field** carries the name of the wireless network. It has a length of 32 bytes. The advantage of using this field is its relatively easy way of manipulation. Most commercial APs provide a method for setting the SSID field. Clients running different operating systems can get the SSID from a beacon directly in user-space, without requiring specially-adapted drivers for the wireless card. The disadvantage is that beacons carrying useful information would overlap with other beacons sent from access points not participating in our solution. Clients might be presented with information about fake networks from non-participating APs. The **BSSID field** (6 bytes long) is the unique identifier of the AP. This overcomes the limitation of using the SSID field, but the small size is not practical for our application. Finally, the **Vendor Specific field** can also be used. The 802.11 standard allows AP vendors to add 253 bytes of information at the end of a beacon frame. The use of this field requires updating the AP's operating system and network stack to add vendor-specific information dynamically in the beacon. The wireless network card driver on the client's side must also be updated to allow reading this field and passing the information to user-space applications. However, vendor specific extensions for reading the fields are currently being developed for modern mobile operating systems (Android, iOS) (Berg, 2009).

The alternatives are not enough to transport enough data in the wireless beacon. Therefore, we further considered solutions to send larger chunks of data. We turn to splitting the messages in smaller data chunks that can fit inside a beacon frame. Each frame would then contain a fragment of the message. Each fragment receives an index. The first and last fragments also contain special delimiters for the limits of the message. The AP then keeps sending the message in a loop. If a client happens to lose the beacon message because of the outside interference, it simply notices that the received fragments are not sequential. In this case it can either discard the fragments or, if possible, reconstitute the message from a second loop of the same message.

In our experiments we considered the Vendor Specific field of being capable of carrying the information. It is the one that offers an acceptable bandwidth for fast information transmission. From the 253 bytes available in this field, we use 3 bytes for control and chunk ordering and the other 250 for transmitting the data. The proposed structure of this custom Vendor Specific field is presented in Figure 2.

| 2B | 1B | 250B |
|---|---|---|
| Seq. No. | First/Last Tag | Data |

Figures 2: Vendor Specific fields.

The *Sequence Number* field is two bytes in length and contains the index inside the message for the currently transmitted data fragment. If we were to use only one byte for this field, our message size would have been limited to 256 fragments of 250 bytes each, leaving us with a maximum message size of 62.5 KB.

The *First/Last Tag* field contains *1* for the first fragment of the message and *2* for the last fragment. All other fragments set this field to *0*. Finally, the *Data* field contains the custom data.

An AP periodically sends beacon frames to make its presence known over the wireless network. Under normal operation these frames carry only the communication parameters advertised by the AP. The default interval between beacons for most APs is set to 100 ms. The time interval offers a good balance between power consumption and responsiveness when scanning the medium for discovering the AP. The interval can be manually set to any value between 1 ms and 65535 ms. In our experiments we considered a beacon interval of 10 ms.

The wireless transmission range depends on the 802.11 version and the AP's producer. Several companies claim outdoors ranges of up to 150 m using the 802.11g protocol. However, we used a 90m range, based on previous research results (Singh, *et al*, 2002). Beyond this limit all packets are assumed lost. Nonetheless, the declared maximum range is that at which the AP is designed to be able to maintain a relatively stable connection with a client. However, we do not need establishing a connection, so the real distance at which beacons are received might be greater, even though the loss ratio increases dramatically with the distance.

Based on the assumed figures, we evaluated the maximum amount of information that can be sent. We used a simple model, with an AP on a road, and several cars (see Figure 3).

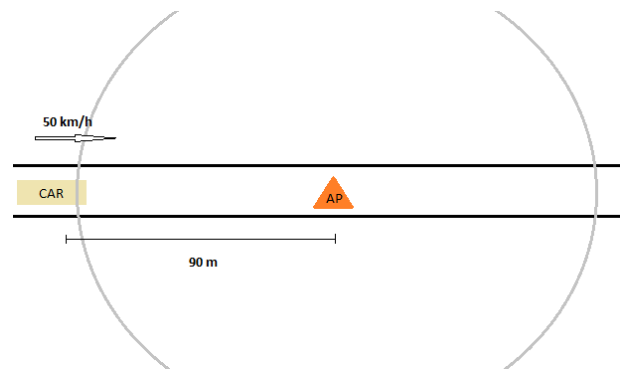

Figures 3: Car approaching an AP.

Considering a speed limit of 50 km/h for a vehicle travelling in the city, the maximum time it takes before it reaches the access point is equal to $90m/(50km/h) = 6.5s$.

A car would have to receive the message in less than *6.5* seconds, before reaching the AP's position. Considering a beacon interval of *10* ms, the AP is able to send *100* packets every second. In this case the vehicle can receive maximum *650* packets before reaching the AP's location. By adding *250* bytes of data in every packet, the amount of information is $650\ packages \cdot 250\ bytes/package = 158\ KB$. So, the maximum amount of data that can be transmitted before the car reaches the AP is *158* KB, and the total amount of data the car is able to receive before completely exiting the transmission range of the AP is *316* KB.

## 3. IMPLEMENTATION DETAILS

We evaluated the mobile advertisement solution using modeling and simulation. We implemented the solution as an extension on top of a realistic traffic simulator called VNSim (Gainaru, *et al*, 2009).

VNSim is a generic VANET traffic simulator incorporating microscopic and macroscopic traffic, mobility and networking models to accurately evaluate the performance of a wide range of VANET technologies. It is designed as a realistic simulator for evaluating the performances of a wide-range of VANET technologies, ranging from wireless networking protocols and dissemination strategies to applications being developed over VANETs.

The simulator is implemented in Java and uses a synthetic mobility model that integrates both microscopic and macroscopic motion. Its mobility model is in charge of importing the map topology and building the dynamic of all vehicles. In addition, the simulator uses a wireless networking model, responsable with the simulation of the networking components and the communication protocols envisioned by a VANET system (see Figure 4).

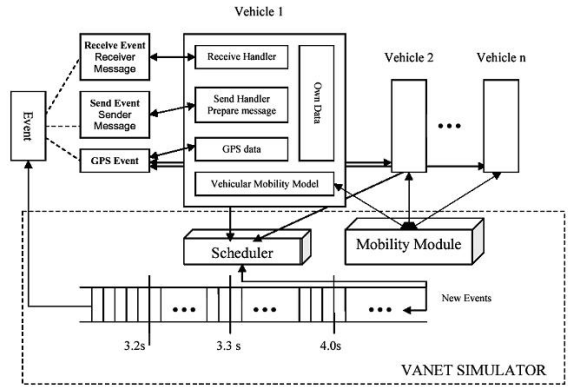

Figures 4: The architecture of VNSim.

For evaluating the proposed Mobile Advertising solution, we extended VNSim with new components and capabilities. On top of the mobility and network models we added new components to simulate behavior and characteristic of APs and cars. Figure 5 illustrates how the mobile advertisement components are integrated with the VNSim simulator. For example, the engine of the simulator maintains a list of cars. We extended it also hold the APs involved in the simulation experiment. Both cars and APs are extension of the same base class CarRunningVITP (Dikaiakos, et al, 2005). The engine creates an instance of the car console and an instance of the statistics class. All the cars and the APs populate these classes directly. At the end of the simulation, the engine displays the statistics.

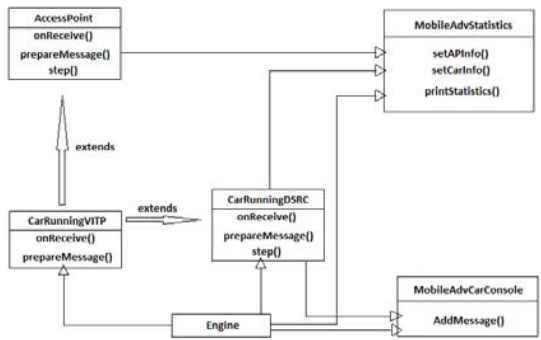

Figures 5: Implementation details.

The extended car receives beacon messages and handles them accordingly. In addition, we introduced a new type of message for caring mobile advertisement (the simulation of the proposed beacon message). The simulated car stores the received messages for a certain period of time. Each car maintains a hash map with the keys being the unique identifiers of the APs sending the messages and a structure holding the received information. When a car receives a beacon frame, it extracts the useful data, and, if it is a new fragment, it further stores it.

Another component added in the simulation is an access point that sends message beacons. The position of the AP on the digital map is set by the user using the graphical user interface. The AP is initialized with default values set for beacon interval, message size and range, and it generates a random message to be sent. The user can specify various parameters for the AP, such as wireless transmission range, size of the transmitted messages, etc.

To simulate a noisy environment we also added a function for randomly dropping packets sent by the AP. The user can specify a probability for the loss of packets.

The output of the simulation experiment consists in the average time required to send a message, the number of message loops needed for a car to receive a complete information in a noisy environment, and the ratio of dropped and received messages.

For the output we added a component that logs and displays statistical information. The statistics contain information about the APs (range, beacon intervals, message size, frames sent, complete loops and time running), about individual cars (completed messages, dropped messages, received frames, duplicate frames and lost frames) and general messaging information (total frames sent, total completed messages, frames received per car, frames lost per car etc.).

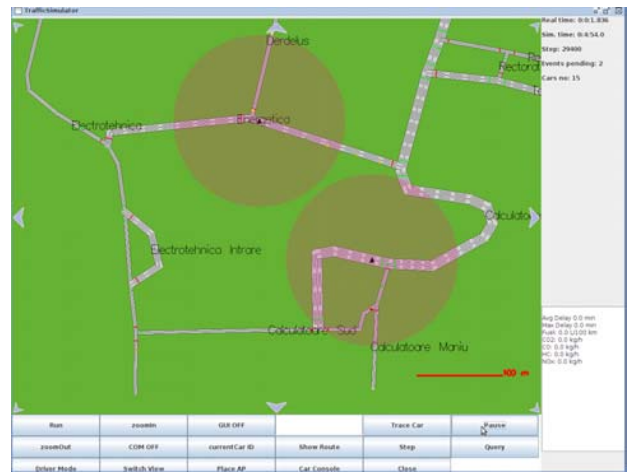

Figures 6: The extended graphical user interface presenting a running experiment.

We also added additional functionality to the graphical output of VNSim (Figure 6). We extended the interactive simulation mode and included a visual representation of the access points and other components used by the mobile advertisement application.

## 4. EXPERIMENTAL RESULTS

We present evaluation experiments designed to determine the maximum throughput of the application. The scenario assumes an AP being placed on an open road, without any intersection nearby and without congested traffic. This represents ideal driving conditions in which cars move with relatively constant speed, without having to slow down or stop. The alternative would be to obtain artificially improved results using reduced car speeds, but this would extend the time cars are within wireless range of the AP.

In the experiments the wireless transmission range of the AP was set to 90 m (and remained constant through the simulation experiment). The beacon interval was set to 10 ms. In an experiment we assumed a 5% packet loss probability, and in another one the target was increased to 10%. We also sent messages with sizes varying between 16KB and 112KB.

Under these conditions, we executed several experiments using increasing sizes for the messages to be transmitted. For each size we computed a "*Message Loss Percentage*", which indicates the percentage of all cars that did not successfully receive the message. The cars moved with an average speed in the interval 60-70 km/h.

Table 1. Evaluation results.

| Packet Loss Probability | Message Size | Message Loss |
|---|---|---|
| 5% | 16 KB | 0% |
| 5% | 32 KB | 0% |
| 5% | 48 KB | 3% |
| 5% | 64 KB | 29% |
| 5% | 80 KB | 58 % |
| 5% | 96 KB | 75% |
| 5% | 112 KB | 90% |
| 10 % | 16 KB | 0% |
| 10 % | 32 KB | 0% |
| 10 % | 48 KB | 8 % |
| 10 % | 64 KB | 58 % |
| 10 % | 80 KB | 80 % |
| 10 % | 96 KB | 93 % |
| 10 % | 112 KB | 100 % |

During these experiments we noticed that for messages smaller than 32 KB the message delivery percentage was 100%, even in a transmission medium with only 10% packet loss probability. For messages exceeding 64KB, the ratio becomes unsatisfactory for our application.

Figure 7 presents the results for the *Message Loss*. The ascending slope of the message loss percentage becomes clearer. We notice that, when increasing the medium packet loss probability, not only do performances drop, but they drop at a faster rate.

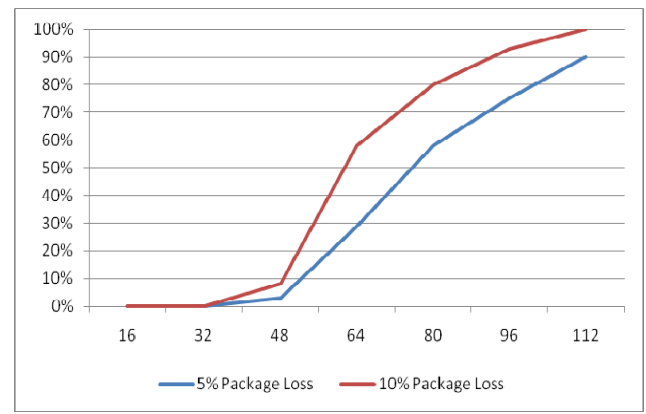

Figures 7: Evaluation results.

For message sizes of approximately 100 KB the messages are almost completely lost, especially in the case of a medium with 10% packet loss probability. The explanation for such a sudden drop in performance is found when examining the time required for a complete message to be sent and the time that a car running at 60-70 Km/h spends in the access point range.

To exemplify the results, let's consider a message of 112 KB. The number of frames required to send the message (considering that one frame can hold up to 250 Bytes) is 459 frames (112 KB / 250 B). The beacon interval was set to 10 ms. Therefore, sending 459 frames would take 4.59 seconds (459 * 10 ms). The time it takes for a vehicle moving at 70 Km/h to traverse the wireless transmission range of an AP (90 m) is 4.6 seconds (90 m / 70 Km/h). Therefore, it takes a vehicle *4.6 seconds* to reach the AP, and it takes a message *4.59 seconds* to be completely sent. When the medium has a 10% packet loss probability chances are that the message needs more than one loop to be successfully transmitted, but during the second transmission the car leaves the access point range.

## 5. RELATED WORK

The world-wide expansion of mobile networks placed mobile technology into the hands of millions of people. Innovative programs have been put in place to disseminate crucial health, social and political data over mobile devices and to use them to collect eyewitness reports and personal health information. Mobile phones can communicate across a variety of online platforms and networks, with high-end phones replicating the capabilities of desktop computers. The primary advantage of mobile data dissemination is in using handheld technology to directly reach the intended recipients.

The problem of data dissemination was approached in many cases for the particular case of Vehicular Ad Hoc Networks. Numerous local incidents occur on road networks daily, many of which may lead to congestion and safety hazards. If vehicles can be provided with information about such incidents or traffic conditions in advance, the quality of driving can be improved significantly in terms of time, distance, and safety. An analysis of several solutions is presented in (Sutariya & Pradhan, 2010).

AdTorrent, an integrated system for search, ranking and content delivery in car networks, previously proposed the notion of Digital Billboards, a scalable "push" model architecture for ad content delivery (Nandan, *et al*, 2006). The mobility model for the urban, vehicular scenario can be used in conjunction with the analytical model for estimating query hit ratio by a system designer to determine the scope of the query flooding as a function of the available storage per vehicle for their application.

Zhao *et al* (2007) propose a solution which considers information source (data center) to disseminate data to many vehicles on the roads. It is noted that periodically pouring data on the road is necessary since vehicles receiving the data may move away quickly, and vehicles coming later still need the data. Caliskan *et al* (2006) focuses on the decentralized discovery of parking places. The proposed model consists of communication between vehicles and fixed infrastructures named as parking automat and also between vehicles.

Unlike previous solution, we propose an approach that takes advantage of short-range wireless network communication. As far as we know it is an idea different from previous solutions by at least: the ability to transfer data without establishing a connection beforehand and the fact that the information is confined to the geographical area in which it could be useful.

## 6. CONCLUSIONS

Mobile advertisement is an application developed on top of a wireless communication infrastructure that provides vehicles on the road with advertisement information related to their current location. The application can be useful to business companies because they can dynamically advertise their products and offers to more people driving their car. We presented a solution for implementing such a system, based on the idea of using access points as emitters for transmitting messages to wireless-enabled devices equipped on vehicles.

Such a solution provides a practical and inexpensive approach for delivering information to vehicles with wireless devices. We presented the approach taken for disseminating information using the beacon frames available in the 802.11 protocol suite. We also presented details about the implementation of the mobile advertisement components on top. We showed how the simulator can be used for evaluating such applications and showed how the mobile advertisement extended its functionality.

The results of the performed experiments reveal that the mobile advertisement solution can transfer a significant amount of data even in difficult conditions, in which cars are moving at increased speeds, or when a congested Wi-Fi network causes significant packet loss. The evaluation results prove that people can more easily disseminate information and can easily discover data of interest using the infrastructure provided by a VANET environment. The performance does not have a large negatively affect on other communications in VANET.

We are currently working on an implementation of the presented solution, using the Vendor Specific field of the 802.11 protocol suite. The solution will be further evaluated in urban traffic scenarios, but we also plan to extend its applications to in-door context-aware dissemination approaches.

## ACKNOWLEDGMENTS

The research presented in this paper is supported by national project: "TRANSYS – Models and Techniques for Traffic Optimizing in Urban Environments", Contract No. 4/28.07.2010, Project CNCSIS-PN-II-RU-PD ID: 238. The work has been co-funded by the Sectoral Operational Programme Human Resources Development 2007-2013 of the Romanian Ministry of Labour, Family and Social Protection through the Financial Agreement POSDRU/89/1.5/S/62557.

## REFERENCES

Berg, J.M. 2009. 'WiFi Overview'. Last accessed March 02, 2011, from http://linuxwireless.org.

Dikaiakos, M.D., S. Iqbal, T. Nadeem, and L. Iftode. 2005, 'VITP: an information transfer protocol for vehicular computing', *Vehicular Ad Hoc Networks 2005*, pp. 30-39.

Dkesson, K.-P., and A. Nilsson. 2002. 'Designing Leisure Applications for the Mundane Car-Commute', *Personal Ubiquitous Comput.*, 6, 3 (January 2002), pp. 176-187.

Gainaru, A., C. Dobre, and V. Cristea. 2009. 'A Realistic Mobility Model based on Social Networks for the Simulation of VANETs', in *Proc. of the VTC-Spring 2009 Conference*, Barcelona, Spain, pp. 1-5.

Lee, W.-H., S.-S. Tseng, and C.-H. Wang. 2008. 'Design and implementation of electronic toll collection system based on vehicle positioning system techniques', *Comput. Commun,* 31, 12 (July 2008), pp. 2925-2933.

O'Grady, M. J., and G.M.P. O'Hare. 2004. 'Just-in-Time Multimedia Distribution in a Mobile Computing Environment'. *IEEE MultiMedia* 11, 4 (October 2004), pp. 62-74.

Sgora, A., D. D. Vergados, P. Chatzimisios. 2009. 'IEEE 802.11s Wireless Mesh Networks: Challenges and Perspectives'. In *Proceedings of MOBILIGHT'2009*. pp.263-271.

Singh, J. P., N. Bambos, B. Srinivasan, D. Clawin. 2002. 'Wireless LAN Performance Under Varied Stress Conditions in Vehicular Traffic Scenarios'. In *Proceedings of Vehicular Technology Conference* (*VTC 2002 – Fall*). pp. 743-747.

Tarnoff, P.J., D.M. Bullock, and S.E. Young. 2009. 'Continuing Evolution of Travel Time Data Information Collection and Processing'.T*ransportation Research Board Annual Meeting 2009 Paper* #09-2030 TRB 88th Annual Meeting Compendium.

Wong, S.H.Y., H. Yang, H. Songwu Lu, and V. Bharghavan. 2006. 'Robust rate adaptation for 802.11 wireless networks', In *Proceedings of the 12th annual international conference on Mobile computing and networking* (MobiCom '06). ACM, New York, NY, USA, pp. 146-157.

Sutariya, D., S. N. Pradhan. 2010. 'Data Dissemination Techniques in Vehicular Ad Hoc Network'. *Intl. Journal of Computer Applications* (0975 – 8887). Volume 8– No.10, pp. 35-39.

Nandan, A., S. Tewari, S. Das, L. Kleinrock. 2006. 'Modeling epidemic query dissemination in Adtorrent Network'. In *Proc. of IEEE CCNC*, pp. 1173-1177.

Zhao, J., Y. Zhang, G. Cao. 2007. 'Data pouring and buffering on the road: A new data dissemination paradigm for Vehicular Ad

Hoc Networks'. *IEEE Transactions on Vehicular Technology*, vol. 56(6). pp. 3266-3277.

Caliskan, M., D. Graupner, M. Mauve. 2006. 'Decentralized discovery of free parking places'. In *Proc. of the Intl. Conf. on Mobile Computing and Networking*, pp. 30- 39.

**BIOGRAPHY**

**CIPRIAN DOBRE** PhD, is lecturer with the Computer Science and Engineering Department of the University POLITEHNICA of Bucharest. The main fields of expertise are Grid Computing, Monitoring and Control of Distributed Systems, Modeling and Simulation, Advanced Networking Architectures, Parallel and Distributed Algorithms. Ciprian Dobre is a member of the RoGRID (Romanian GRID) consortium and is involved in a number of national projects (CNCSIS, GridMOSI, MedioGRID, PEGAF) and international projects (MonALISA, MONARC, VINCI, VNSim, EGEE, SEE-GRID, EU-NCIT). His research activities were awarded with the Innovations in Networking Award for Experimental Applications in 2008 by the Corporation for Education Network Initiatives (CENIC).